\documentclass{fundam}

\usepackage{comment}
\usepackage{url} 
\usepackage[ruled,lined]{algorithm2e}
\usepackage{graphicx}
\usepackage{tikz}
\usetikzlibrary{automata, positioning, arrows}

\usepackage[numbers]{natbib}
\usepackage{graphicx}
\usepackage{amssymb}
\usepackage{amsmath}
\usepackage{tabularx}
\usepackage{lscape}
\usepackage[none]{hyphenat}
\usepackage{float}
\usepackage{booktabs}
\usepackage{subcaption}
\usepackage[none]{hyphenat}
\usepackage{rotating}
\usepackage[misc,geometry]{ifsym}

\begin{document}
	\sloppy


\setcounter{page}{285}
\publyear{22}
\papernumber{2111}
\volume{185}
\issue{4}

  \finalVersionForARXIV


	\title{On Random Number Generation for Kernel Applications}

	\author{Kunal Abhishek\thanks{Address of correspondence: Society for Electronic Transactions and Security (SETS),
                        MGR Knowledge City, C.I.T. Campus, Taramani, Chennai,  600113 - India.\newline \newline
          \vspace*{-6mm}{\scriptsize{Received June 2021; \ accepted April  2022.}}}
\\
		Society for Electronic Transactions and Security (SETS) \\
		Chennai, India\\
		kunalabh@gmail.com
		\and E. George Dharma~Prakash~Raj\\
		School of Computer Sciences, Engineering and Applications\\
		Bharathidasan University, Tiruchirappalli, India \\
		georgeprakashraj@yahoo.com}
	
	\maketitle
	
	\runninghead{K. Abhishek and  E.G. Dharma~Prakash~Raj}{On Random Number Generation for Kernel Applications}
	
\vspace*{-5mm}
\begin{abstract}
An operating system kernel uses cryptographically secure pseudorandom number generator (CSPRNG) for creating address space layout randomization (ASLR) offsets to protect memory addresses of processes from exploitation, storing users' passwords securely and creating cryptographic keys.~However, at present, popular kernel CSPRNGs such as Yarrow, Fortuna and /dev/(u)random which are used by MacOS/iOS/FreeBSD, Windows and Linux/Android kernels respectively lack the very crucial property of non-reproducibility of their generated bitstreams which is used to nullify the scope of predicting the bitstream.~This paper proposes a CSPRNG called Cryptographically Secure Pseudorandom Number Generator for Kernel Applications (KCS-PRNG) which generates non-reproducible bitstreams.~The proposed KCS-PRNG presents an efficient design uniquely configured with two new non-standard and verified elliptic curves and clock-controlled Linear Feedback Shift Registers (LFSRs) and a novel method to consistently generate non-reproducible random bitstreams of arbitrary lengths.~The generated bitstreams are statistically indistinguishable from true random bitstreams and provably secure, resilient to important attacks, exhibits backward and forward secrecy, exhibits exponential linear complexity, large period and huge key space.
	\end{abstract}
	

	\section{Introduction}
	\label{intro}
	A random number generator (RNG) is classified in two basic classes \cite{kocc2009cryptographic}: first, a deterministic random number generator (DRNG) or a pseudorandom number generator (PRNG) which needs a seed value as input and produces random looking bitstreams using some deterministic algorithm. Second, a true random number generator (TRNG) which uses physical and non-physical sources to generate true randomness. It is imperative to note that unlike PRNG or DRNG, TRNG does not need any seed but uses non-deterministic effects or physical experiments to generate the true random bits \cite{kocc2009cryptographic}. The significant differences between TRNG and PRNG are that TRNG generates non-reproducible arbitrary length random bitstreams without using any seed or initializer whereas the PRNG generates arbitrary length pseudorandom bitstreams using a seed value or initializer. TRNG is slow, having infinite period, costly in deployment and has the possibility of manipulation. Unlike TRNG, PRNG has less development and deployment cost (no need of dedicated hardware) but can produce reasonably good random looking bitstreams.
	
	The design goals of RNG heavily depend on its target applications. A simple application like stochastic simulations or Monte Carlo integrations may require RNG to generate nothing more than a random looking bitstream \cite{kocc2009cryptographic}. However, a sensitive application of RNG like an operating system kernel on top of which entire critical systems or applications run, certainly requires RNG to generate high quality pseudorandom bitstreams which are also provably secure, unpredictable and must be non-reproducible.
	
	A kernel uses a RNG to create ASLR offsets \cite{marco2019address}, generate salts to securely store users passwords \cite{Tanenbaum2006} and generate random keys to implement various cryptographic primitives such as authentication etc. The ASLR is one of the most important techniques used by the kernel (in special cases termed as Kernel-ASLR or KASLR) which randomizes the process layout to protect the locations of the targeted functions such as stack, heap, executable, dynamic linker/loader etc. \cite{marco2019address}. The ASLR not only demands statistically qualified high quality pseudorandom number generator but also requires the output bitstream to be provably secure and unpredictable. Hence, a CSPRNG (or simply a PRNG with regular entropy inputs for unpredictability) is a preferred type of RNG for kernel applications. There are many good CSPRNGs which are implemented in various operating systems and are used by their kernels. Fortuna, Yarrow and /dev/(u)random are the popular CSPRNGs which are currently implemented by Windows, MacOs/iOS/FreeBSD and Linux/Android operating systems respectively \cite{dodis2017eat,dorre2015pseudo}. In this paper, a new CSPRNG which exhibits `non-reproducibility' property of a TRNG is proposed taking security of the above kernel applications into consideration.

\medskip	
	In particular, the key contributions of this paper are as follows:
	\begin{itemize}
\itemsep=0.8pt
		\item{A novel CSPRNG design comprises of two non-standard and verified secure elliptic curves and nine LFSRs uniquely configured in a clock-controlled fashion to attain exponential linear complexity is used to construct the proposed KCS-PRNG.}
		\item{A novel architecture of the KCS-PRNG is proposed to mitigate the gap of `non-reproducibility' property.}
		\item{Two new non-standard and verified elliptic curves are introduced in this paper which are used by the proposed KCS-PRNG to mitigate the gap of `non-reproducibility' property. Both elliptic curves are generated randomly over 256-bit prime fields to ensure cryptographic and implementation security.}
		\item{Extensive security analysis of the proposed KCS-PRNG is carried out to ensure theoretical security.}
		\item{Experimental validation and demonstration of statistical qualities of randomness using National Institute of Standards and Technology (NIST), Diehard, TestU01 test suites.}
		\item{Experimental validation and demonstration of `non-reproducibility' property of the proposed KCS-PRNG.}
		\item{The proposed KCS-PRNG is compared with present kernel CSPRNGs such as Fortuna, Yarrow and dev/random and an existing PRNG \cite{alhadawi2019designing}. The KCS-PRNG is also compared with an existing TRNG \cite{anandakumar2019fpga} in context of non-reproducibility of the generated random bitstreams.}
	\end{itemize}
	Rest of the paper is organized as follows: Section $\ref{sec2}$ briefly discusses the randomness requirements of the kernel applications and standard RNG requirements. Section $\ref{sec3}$ reviews current CSPRNGs implemented by popular operating system kernels. Section $\ref{sec4}$ presents the proposed design of the KCS-PRNG. Subsequently, Section $\ref{sec5}$ presents the security analysis and Section $\ref{sec6}$ elaborates experimental validation and demonstration of the proposed KCS-PRNG. Section $\ref{sec7}$ presents the details of the two new elliptic curves selected over large prime fields for use in the proposed KCS-PRNG.~Section $\ref{sec8}$ shows the important obtained results of the proposed KCS-PRNG.~Section $\ref{sec9}$ briefly analyses the performance of KCS-PRNG. Section $\ref{sec10}$ compares KCS-PRNG with existing kernel CSPRNGs as well as recent PRNG, CSPRNG and TRNG used by various cryptographic applications. Finally, Section $\ref{sec11}$ concludes the findings of this paper.
	
	\section{Preliminaries}\label{sec2}
	\subsection{Randomness for Kernel Applications}	\label{subsec1}
	
\begin{figure}
		\centering
		\includegraphics[width=2.8in]{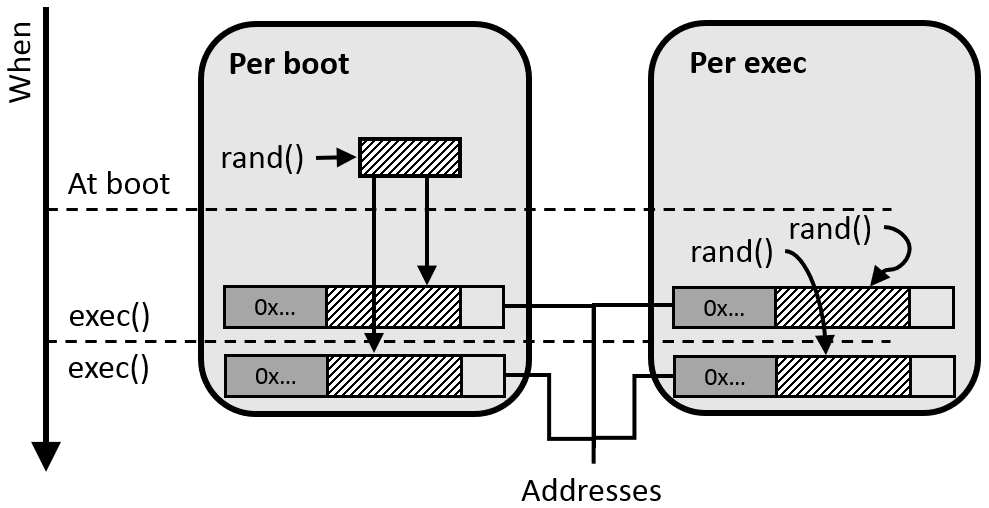}
		\caption{Per-boot versus Per-exec randomization \cite{marco2019address}}\label{fig_aslr}
	\end{figure}

\begin{figure}
		\centering
		\includegraphics[width=2.8in]{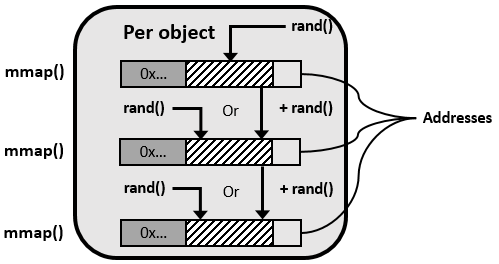}
		\caption{Per-object randomization \cite{marco2019address}}\label{fig_PerObject}
	\end{figure}

	One of the most important kernel applications that requires high quality randomness is ASLR \cite{marco2019address} which is an efficient mitigation technique against remote code execution attacks by randomizing the memory address of processes to disable memory exploitation. The ASLR currently uses CSPRNG to randomize the logical elements contained in the memory objects at the time of pre-linking (at the time of installation of the application), per-boot (on every time the system boots), per-exec (when new executable image is loaded in memory called pre-process randomization), per-fork (every time a new process is created) and per-object (every time a new object is created). Figure $\ref{fig_aslr}$ shows the Per-boot versus Per-exec randomization to point out when randomization takes place in both the per-boot and per-exec processes. Similarly, Figure $\ref{fig_PerObject}$ shows that $mmap()$ system call allocates all the objects side by side in the $mmap\_area$ area during the per-object randomization taking place. The $rand()$ provides random bits of desired length to the objects as shown in Figure  $\ref{fig_PerObject}$.

	Moreover, the degree of security provided by ASLR technique depends on the predictability of the random memory layout of a program and therefore, `non-reproducibility' of the random sequences used in ASLR needs to be additonally ensured. This particular requirement is also attended in the present work.
	
	Another important kernel application is the Morris-Thompson scheme \cite{Tanenbaum2006,silberschatz2015instructor} which associates a $n$-bit random number with each password and concatenate and then encrypt together before storing it in the password file. A CSPRNG is used whenever a password is changed and a random number is required.
	
	\subsection{RNG requirements}	\label{subsec4}
	Koc \cite{kocc2009cryptographic} and Schneier \cite{schneier2007applied} collated the properties that various classes of RNG exhibit and formulated the following requirements:
	\begin{enumerate}
\itemsep=0.9pt
		\item	$R1$ \label{R1}: A random sequence generated by a RNG should have good statistical properties.
		
		This requirement enables a RNG with a large period.
		\item	$R2$ \label{R2}: A random sequence generated by a RNG should be unpredictable.
		
		This requirement makes the prediction of the next bit infeasible in the stream, given the complete knowledge of the algorithm or hardware which generates the sequence and all of the previous bits in the stream. This gives the notion of Backward Secrecy.
		\item	$R3$ \label{R3}: A random sequence generated by a RNG should not allow to compute previous internal state or values of the generator even if the internal state is known.
		
		This gives the notion of Forward Secrecy.
		\item $R4$ \label{R4}: A random sequence generated by a RNG should not be reliably reproduced.
		
		If the RNG is run twice with exactly the same input, it should produce two completely unrelated random sequences.
	\end{enumerate}
	From definition, a PRNG meets only $R\ref{R1}$ requirement whereas CSPRNG meets $R\ref{R1}$, $R\ref{R2}$ and $R\ref{R3}$ requirements of RNG \cite{schneier2007applied}. However, a TRNG meets $R\ref{R2}$, $R\ref{R3}$ and $R\ref{R4}$ requirements of the RNG \cite{schneier2007applied}. In this paper, the proposed KCS-PRNG is designed in such a way that it meets the $R\ref{R1}$, $R\ref{R2}$ and $R\ref{R3}$ requirements along with the $R\ref{R4}$ requirement of RNG to a practical extent.
	
	\section{Cryptographically secure random number generators for kernels} \label{sec3}

	Linux and Android kernels use /dev/random and /dev/urandom which are considered as CSPRNG i.e. the PRNG with inputs (meeting the requirement $R\ref{R2}$) for randomness generation. The limitations of these CSPRNGs are that they do not have enough entropy in the pool and they are not generating keys larger than the hash function that they used internally \cite{viega2003practical}. /dev/random keeps awaiting for the entropy pool to get sufficiently filled in, which results in diminished performance of the generator. /dev/random meets the RNG requirements $R\ref{R1}$, $R\ref{R2}$ and $R\ref{R3}$ but does not meet the $R\ref{R4}$ requirement.~Though /dev/urandom has provision for unblocked fast supply of random sequences through unblocking pool of entropy but faces predictability issues \cite{dodis2013security}. /dev/urandom meets the requirements $R\ref{R1}$ and $R\ref{R3}$ but does not meet the requirement $R\ref{R2}$ and $R\ref{R4}$.
	
	Yarrow \cite{kelsey1999yarrow} is a PRNG with true random inputs used by MacOS/iOS/FreeBSD kernels. This CSPRNG is too complex and under-specified in entropy handling context and also slow to provide an initial seed \cite{viega2003practical}. It uses Triple Data Encryption Standard (DES) block cipher for pseudorandom bitstream generation. Like /dev/random, Yarrow meets the requirements $R\ref{R1}$, $R\ref{R2}$ and $R\ref{R3}$ but does not meet the requirement $R\ref{R4}$.
	
	Fortuna \cite{ferguson2011cryptography} is a popular CSPRNG and a refinement over Yarrow, used by the Windows kernel which uses its entropy effectively. It uses Advanced Encryption Standard (AES)-like cipher for the generator with 256-bit size of the block cipher key and a 128-bit counter. Fortuna produces a very good throughput of 20 clock cycles per byte on CPU type PC \cite{ferguson2011cryptography} and 7.2 Mbps throughput in software \cite{mcevoy2006fortuna}. Fortuna implicitly accumulates entropy through hash, partitions the incoming entropy into multiple entropy pools and uses its pools at different rate for output generation in order to guarantee that at least one pool will remain available for use \cite{dodis2017eat}. Though Viega \cite{viega2003practical} observed that it completely foregoes the entropy estimation and Fortuna and Yarrow both do not exhibit information-theoretic security. Like Yarrow, Fortuna also meets the requirements $R\ref{R1}$, $R\ref{R2}$ and $R\ref{R3}$ but does not meet `non-reproducibility' i.e., the requirement $R\ref{R4}$.
	
	It is imperative to note that the present kernel CSPRNGs do not meet the requirement of `non-reproducibility' i.e., the requirement $R\ref{R4}$ which is a crucial feature that helps to prevent the kernel better from exploitation as discussed in Section $\ref{subsec1}$. In this work, the proposed KCS-PRNG is designed in such a way that all the four requirements ($R\ref{R1}$ to $R\ref{R4}$) of an ideal RNG are met to ensure better prevention of the kernel from exploitation.
	
	\section{The proposed design of KCS-PRNG} \label{sec4}

	Generation of high quality cryptographically secure pseudorandom bitstreams is an intricate task which needs efficient design of the generator taking statistical properties of randomness ($R\ref{R1}$), unpredictability ($R\ref{R2}$, $R\ref{R3}$) and non-reproducibility ($R\ref{R4}$) of the output bitstreams into consideration.~For this reason, the proposed KCS-PRNG binds two modules in its design: first, a combination of two cryptographically safe elliptic curves and a nonlinear Sequence Generator consisting of nine clock-controlled LFSRs in alternating step configuration.~Following are the design decisions and assumptions of the proposed KCS-PRNG:
	
\subsection{Selection of elliptic curves} \label{SelECs}

	The main motivation of using elliptic curves in the proposed KCS-PRNG instead of stream ciphers/block ciphers like ChaCha20  and Triple DES or AES respectively as used by /dev/(u)random \cite{viega2003practical}, Yarrow \cite{kelsey1999yarrow} and Fortuna \cite{ferguson2011cryptography} respectively is that one can choose different points on the selected elliptic curve to generate completely unrelated bitstreams under identical start conditions.~Hence, the combination of elliptic curves and clock-controlled LFSRs in the proposed KCS-PRNG generates non-reproducible cryptographically secure pseudorandom bitstreams.~Moreover, the combination of elliptic curve and LFSR has been proven to exhibit enhanced randomness properties \cite{gong2002linear}. Two elliptic curves are used in KCS-PRNG for added complexity where each elliptic curve provides nearly $2^{128}$ key space.~The advantages of keeping elliptic curves with the clock-controlled LFSRs are twofold: first, the elliptic curves are used for mitigating the gap of `non-reproducibility' property ($R\ref{R4}$) by the proposed method of replacing them periodically from a look-up table.~Second, elliptic curves are used to generate bitstreams which are non-invertible due to underlying hard Elliptic Curve Discrete Logarithm Problem (ECDLP) and hence, they make the proposed KCS-PRNG provably secure as well as forward secure to resist backtracking attacks.~However, the choice of elliptic curves is considered to be a randomly generated one rather than the standard elliptic curves with fixed coefficients as being recommended by agencies like NIST \cite{kerry2013digital}, Brainpool \cite{brainpoolbrainpool} etc., so that a look-up table can be created consisting of reasonably large number of cryptographically secure elliptic curves of one's choice.~The random derivation of elliptic curve parameters ensures trust and transparency in the implementation of elliptic curves \cite{bernstein2015manipulate}.~The details of the two elliptic curves selected for use in the KCS-PRNG are presented in Section $\ref{sec7}$.~One can create a look-up table consisting of elliptic curves of 256 bit field order of one's choice for use in the KCS-PRNG.~The discussion on generation mechanism of elliptic curves is outside the scope of this paper due to space limitation.
	
\subsection{Selection of clock-controlled LFSRs} \label{SelLFSR}

The proposed KCS-PRNG is targeted for integration in the operating system kernel and therefore, it is implemented in software. However, implementation of LFSR in software is slower than its hardware implementation \cite{schneier2007applied,mukhopadhyay2006application}. To address this performance issue, the Galois scheme is selected for optimal performance gain by the LFSRs in software without compromising the LFSR period and its cryptographic properties \cite{schneier2007applied}. The chosen Galois configuration also saves excess operations as all the XOR operations are performed as a single operation \cite{schneier2007applied}. A nonlinear Sequence Generator consisting of nine LFSRs $L_1, L_2, L_3, L_4, L_5, L_6, L_7, L_8$ and $L_9$ with corresponding primitive polynomial degrees 29, 31, 37, 41, 43, 47, 53, 59 and 61 respectively is designed. The primitive polynomials for these LFSRs feedback functions are 
	\begin{center}
		$L_1 = x^{29}+x^{25}+x^{21}+x^{17}+x^{14}+x^{10}+x^{6}+x^3+1$, \\
		$L_2 = x^{31}+x^{27}+x^{23}+x^{19}+x^{15}+x^{11}+x^{7}+x^3+1$, \\
		$L_3 = x^{37}+x^{32}+x^{27}+x^{23}+x^{18}+x^{13}+x^{9}+x^5+1$, \\
		$L_4 = x^{41}+x^{36}+x^{31}+x^{26}+x^{20}+x^{15}+x^{10}+x^5+1$, \\
		$L_5 = x^{43}+x^{37}+x^{31}+x^{25}+x^{20}+x^{15}+x^{10}+x^5+1$, \\
		$L_6 = x^{47}+x^{41}+x^{35}+x^{29}+x^{23}+x^{17}+x^{11}+x^5+1$, \\
		$L_7 = x^{53}+x^{46}+x^{40}+x^{33}+x^{26}+x^{19}+x^{13}+x^7+1$, \\
		$L_8 = x^{59}+x^{52}+x^{44}+x^{36}+x^{29}+x^{22}+x^{14}+x^7+1$, \\
		$L_9 = x^{61}+x^{53}+x^{45}+x^{38}+x^{30}+x^{23}+x^{15}+x^7+1$.
	\end{center}
	These primitive polynomials used by the nine LFSRs have uniformly distributed feedback coefﬁcients selected from \cite{rajski2003primitive}. These nine LFSRs $L_1, L_2, \cdots, L_9$ are further divided into three groups called Sequence Generator 1 ($SG_1$), Sequence Generator 2 ($SG_2$) and Sequence Generator 3 ($SG_3$). $SG_1$ has three LFSRs $L_1, L_2$ and $L_3$ whose output streams $x_1, x_2$ and $x_3$ are combined nonlinearly using nonlinear function
	\begin{equation} \label{nonLinearFn}
		y_1: f(x_1, x_2, x_3) = x_1x_2 \oplus x_2x_3\oplus x_3x_1 \vspace{1mm}
	\end{equation}
	The resulting sequence $y_1$ has period $(2^{L_1}-1)(2^{L_2}-1)(2^{L_3}-1)$ and linear complexity $(L_1L_2+L_2L_3+L_1L_3)$.~Similarly, the period and linear complexity of the sequence $y_2$ generated from $SG_2$ using $L_4$, $L_5$, $L_6$ are $(2^{L_4}-1)(2^{L_5}-1)(2^{L_6}-1)$ and  $(L_4L_5+L_5L_6+L_6L_4)$ respectively whereas the period and linear complexity of the sequence $y_3$ generated from $SG_3$ using $L_7$, $L_8$, $L_9$ are $(2^{L_7}-1)(2^{L_8}-1)(2^{L_9}-1)$ and $(L_7L_8+L_8L_9+L_9L_7)$ respectively.~It may be noted that the initial state bits of all LFSRs together are $\sum_{i=1}^{9} L_i$ = 401 bits.
	
\begin{figure*}[!h]
		\centering
		\includegraphics[width=5.3in]{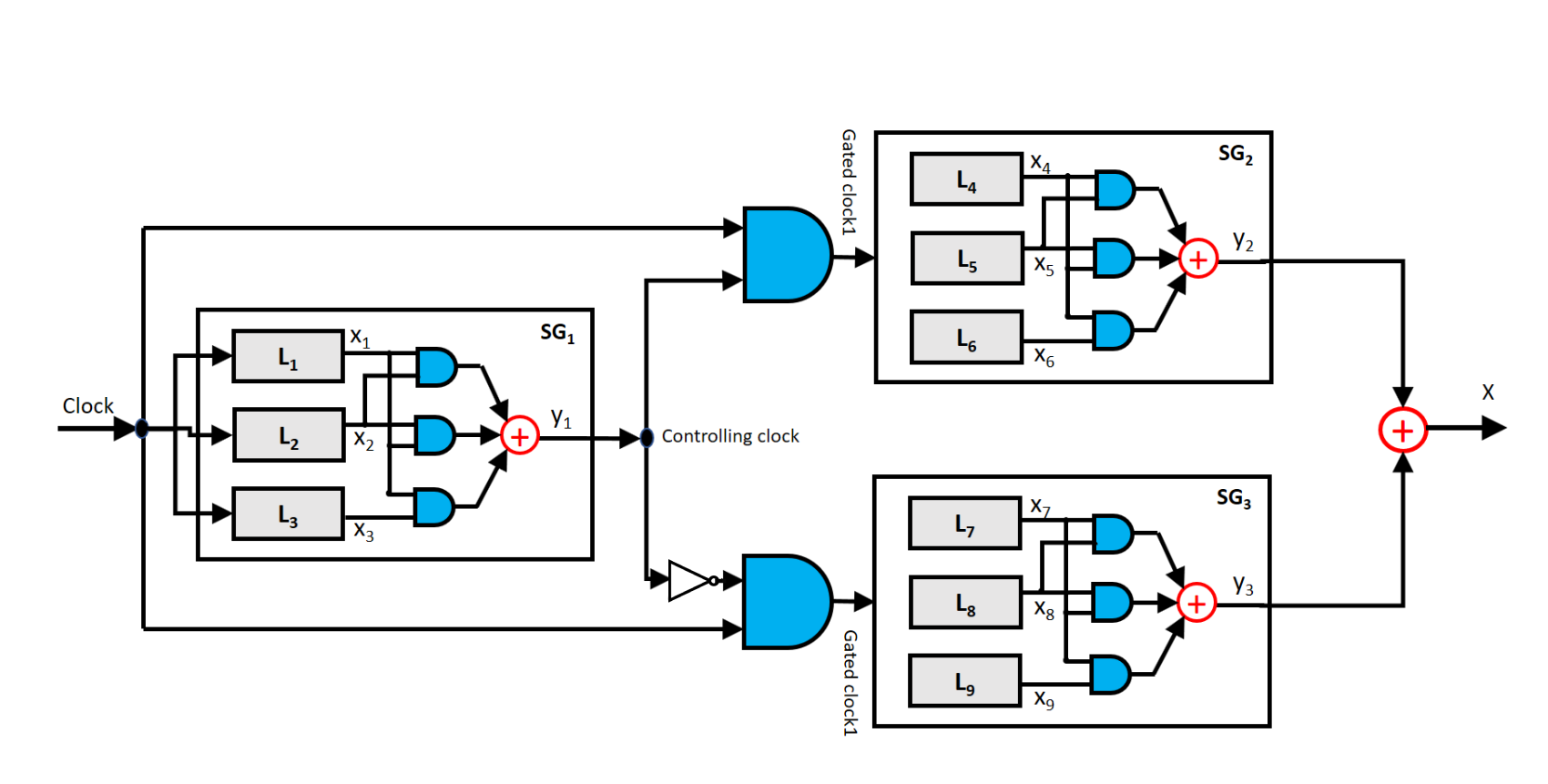}\vspace*{-2mm}
		\caption{Sequence Generator in Alternating Step Configuration}
		\label{fig_AlterStepSeqGen}  \vspace*{5mm}     
	\end{figure*}

\medskip
	$SG_1$, $SG_2$ and $SG_3$ are configured in alternating step scheme to provide high linear complexity and large period of the Sequence Generator \cite{menezes2018handbook}. $SG_1$ is considered as the Controller of the Sequence Generator in the alternating step mode. It is known that the linear complexity $LC(x)$ of the overall alternating step generator is bounded as follows \cite{menezes2018handbook}:
	\begin{equation} \label{linComplexity}
		(LC_2 + LC_3)^{2LC_1-1} < LC(x) \leq (LC_2 + LC_3)^{2LC_1}	
	\end{equation}
	where $LC_1, LC_2$ and $LC_3$ are the linear complexities of $SG_1, SG_2$ and $SG_3$ respectively. The Alternating Step Sequence Generator used in the proposed KCS-PRNG is depicted in Figure $\ref{fig_AlterStepSeqGen}$ and described in Algorithm $\ref{seqGen}$ \cite{menezes2018handbook}.
	
	\begin{algorithm}[h!]
		\caption{Alternating Step Sequence Generator using Clock-controlled LFSRs}
		\SetKwInOut{KwIn}{Input}
		\SetKwInOut{KwOut}{Output}
		\KwIn{Sequence Generators $SG_1, SG_2$ and $SG_3$}
		\KwOut{bit length $n$}
		\For{$i \gets 1$ \textbf{to} $n$ }  {
			$SG_1$ is clocked   \\
			\eIf{ $SG_1 == 1$} {
				$SG_2$ is clocked. \tcc{$SG_3$ is not clocked but its previous output bit is repeated.~In case of the first clock cycle, previous output bit of $SG_3$ is taken as 0.}}
			{ $SG_3$ is clocked \tcc{$SG_2$ is not clocked but its previous output bit is repeated.~In~case of the first clock cycle, previous output bit of $SG_2$ is taken as 0.}
			}
			\KwRet{$y_2$ $\oplus$ $y_3$}	\tcp{Output of Sequence Generator in alternating step}
		}\label{seqGen}
	\end{algorithm}
	

	\begin{figure*}[!h]
\vspace*{5mm}
		\centering
		\includegraphics[width=5.2in]{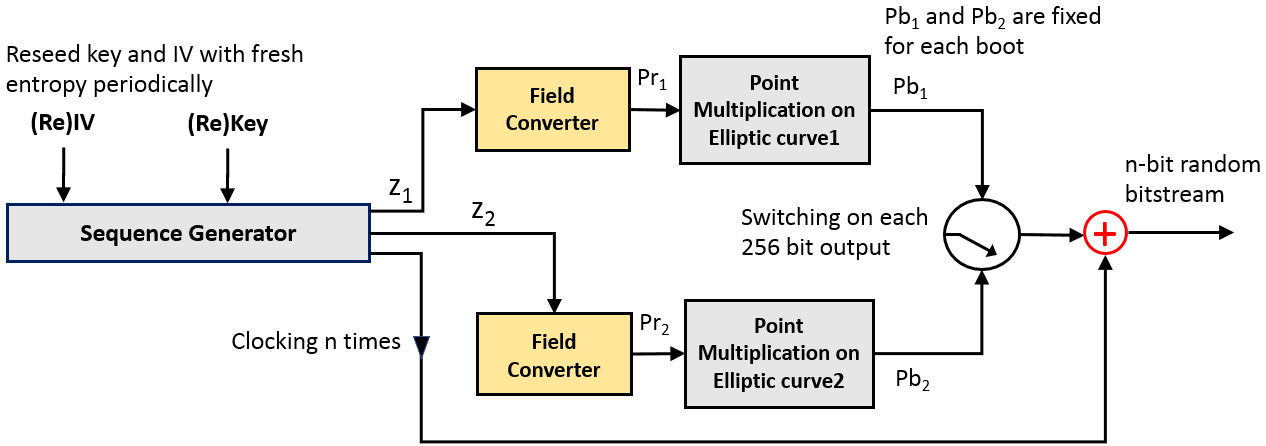}
		\caption{The proposed KCS-PRNG Architecture}
		\label{fig_KCSPRNG}
	\end{figure*}
	
	\subsection{The proposed novel KCS-PRNG architecture} \label{architecture}

	The proposed KCS-PRNG architecture is shown in Figure $\ref{fig_KCSPRNG}$. The KCS-PRNG uses a Field converter, Elliptic curve Point Multiplication and a Selector in addition to the Sequence Generator and two elliptic curves in its design.

	\begin{algorithm}[h!] \caption{Selection of 2 Elliptic curves}
		\SetKwInOut{KwIn}{Input}
		\SetKwInOut{KwOut}{Output}
		
		\KwIn{Look-up Table $\mathcal{T}$($EC_n$, $EC_n\_ID\_Status$) where $n$ is number of elliptic curves}
		\KwOut{Elliptic curves $EC_r$, $EC_s$ from $\mathcal{T}$ where $r, s \in [1,n]$ and $r \neq s$}
		Count $n$	\tcp{Elliptic curves with $EC_n\_ID\_Status=0$ $\forall$ $n$ in $\mathcal{T}$}
			\eIf{$n \geq 2$} {
				Fetch $EC_r$, $EC_s$ from $\mathcal{T}$ where $EC_r\_ID\_Status=0$ and  $EC_s\_ID\_Status=0$  \\
				Set $EC_r\_ID\_Status \longleftarrow 1, EC_s\_ID\_Status \longleftarrow 1$  \\
				Update $\mathcal{T}$ \\
				\KwRet{$EC_r, EC_s$}
			}
			{
				Set $EC_n\_ID\_Status=0$ $\forall$ $n$ in $\mathcal{T}$ \\
				Go to previous step
			}
			\label{ecsSel}
		\end{algorithm}
		
		The two elliptic curves are selected using the procedure as shown in Algorithm $\ref{ecsSel}$. A look-up table $\mathcal{T}$ with tuples ($EC$, $EC\_ID\_Status$) is created where $EC$ is the elliptic curve and $EC\_ID\_Status$ is the flag value to mark 0 for `un-used curve' and 1 for the `used curve'. $\mathcal{T}$ consists of 256 elliptic curves initially which are randomly generated and are cryptographically secure non-standard curves. All elliptic curves in $\mathcal{T}$ are initially marked with $EC\_ID\_Status=0$. On each reboot of the proposed KCS-PRNG, it picks up two elliptic curves from $\mathcal{T}$ using Algorithm $\ref{ecsSel}$ and sets the corresponding $EC\_ID\_Status=1$ of both the used elliptic curves in $\mathcal{T}$. The advantage of $\mathcal{T}$ is that even if the same seed (entropy) is supplied to the proposed KCS-PRNG on reboot of the generator, two new elliptic curves with $EC\_ID\_Status=0$ will be selected from $\mathcal{T}$. The change of elliptic curves on each reboot of the KCS-PRNG changes the final output by altering the masking value between the output bits of the elliptic curves and the Sequence Generator. Hence, entirely unrelated bitstream are obtained as the output of the proposed generator even using exactly the same seed as input. When all elliptic curves in $\mathcal{T}$ are used then $EC\_ID\_Status$ flags are reset to $0$ for all elliptic curves in $\mathcal{T}$ in order to maintain unblocked supply of elliptic curves to the KCS-PRNG. More elliptic curves can be inserted into $\mathcal{T}$ to consistently mitigate the requirement of `non-reproducibility' property  $R\ref{R4}$ of the KCS-PRNG. Here, the mitigating factor of the the RNG requirement $R\ref{R4}$ is directly proportional to the number of un-used elliptic curves available in $\mathcal{T}$. This idea makes the proposed KCS-PRNG to mitigate the RNG requirement $R\ref{R4}$ to a practical extent.
		
\subsection{Initialization of KCS-PRNG} \label{initialization}

\begin{figure}
			\centering
			\includegraphics[width=2.5in]{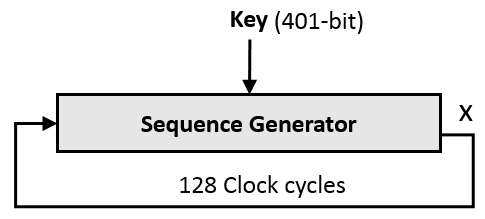}
			\caption{Initialization Stage 1: Loading and diffusion of the key}
			\label{init1}
		\end{figure}

\begin{figure*}
			\centering
			\includegraphics[width=6in]{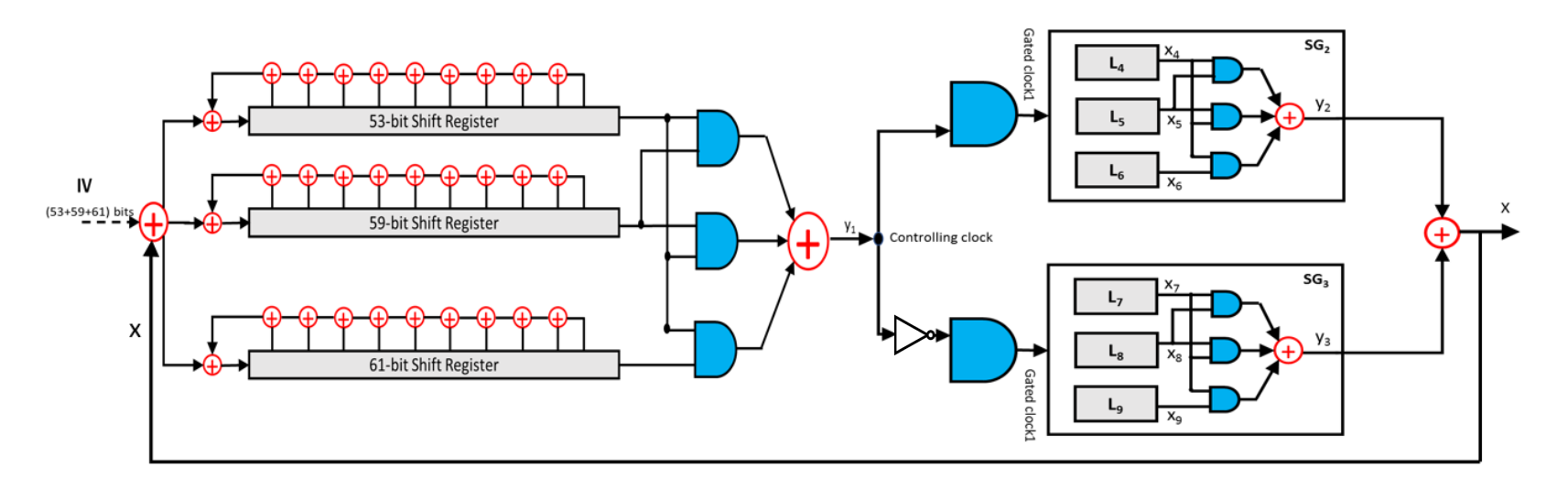}
			\caption{Initialization Stage 2: Loading of IV}
			\label{init2}
		\end{figure*}

The proposed KCS-PRNG uses two phases of pseudorandom bitstreams generation. In the first phase, the Sequence Generator is initialized whereas in the second phase, the desired number of bits of the pseudorandom sequence are generated using the Sequence Generator and the elliptic curves. The initialization phase involves two stages which includes, first, loading the key and initialization vector (IV) into the generator and second, diffusing the key-IV pair across the entire states of the Sequence Generator \cite{teo2013analysis} as shown in Figure $\ref{init1}$ and Figure $\ref{init2}$ described in the Algorithm $\ref{init}$.
		
        \begin{algorithm}[h!] \small
     \caption{Initialization of Sequence Generator}
			\SetKwInOut{KwIn}{Input}
			\SetKwInOut{KwOut}{Output}
			\KwIn{401-bit entropy for Key and 173-bit entropy for Initialization Vector (IV)}
			\KwOut{Initialized Sequence Generator}
			\tcp{Stage 1: Loading LFSRs from the input Key}
			Initialize $SG_1, SG_2$ and $SG_3$ with 401-bit Key \\
			\If{ MSB of any LFSR == 0} {
				Ensure MSB of LFSR as 1
			}
			\tcp{Stage 2: Diffusion of key into all LFSRs states}
			Clock Sequence Generator 128 times
			
			\tcp{Stage 1: Loading 173-bit IV to $SG_1$}
			\For{$i \gets 1$ \textbf{to} $173$} {
				Clock $SG_1$ with feedback = Feedback bit $\oplus$ IV bit $\oplus$ output bit of Sequence Generator
				
			}
			\tcp{Stage 2: Diffusion of IV into all LFSRs states in $SG_1$}
			Clock Sequence Generator 128 times\\
			\If{ MSB of the Sequence Generator == 0} {
				Ensure MSB of the Sequence Generator as 1 }
			\KwRet{Initialized Sequence Generator}
			\label{init}
		\end{algorithm}

\medskip		
Algorithm $\ref{init}$ takes 574-bit of entropy bits which are harvested from various physical non- de\-terministic noise sources and generates 401-bit of key and 173-bit of Initialization Vector (IV). The key is first parallelly loaded into $SG_1, SG_2$ and $SG_3$ of the Sequence Generator.~It is ensured that all the Most Significant Bits (MSBs) of $L_1, L_2$ and $L_3$ will be set to 1. The Sequence Generator is then clocked 128 times so that the key is diffused across the entire states of all the nine LFSRs $L_1, L_2, \cdots, L_9$ and a new state of the Sequence Generator is obtained as shown in Figure $\ref{init1}$. Further, a 173-bit IV is loaded into $L_1, L_2$ and $L_3$ of $SG_1$ in bitwise fashion by XORing with the feedback bit of the LFSR and the output bit of the Sequence Generator as shown in Figure $\ref{init2}$. The Sequence Generator is once again clocked 128 times to diffuse the IV completely among the LFSRs in $SG_1$ and gets entirely new states of all the nine LFSRs. It is ensured that the MSBs of all the nine LFSRs $L_1, L_2, \cdots, L_9$ are set to 1. Finally, the initialized Sequence Generator is returned.
		
\subsection{KCS-PRNG bitstream generation}

	\begin{algorithm}[!b]\small
       \caption{Elliptic curve point multiplication}
			\KwIn{Secrets $P_{r_1}$ and $P_{r_2}$ for 2 elliptic curves}
			\KwOut{Points $P_{b_1}$ and $P_{b_2}$ of 2 elliptic curves in integer form}
				$P_{b_1} \longleftarrow G_1 \times P_{r_1}$  	\tcc{$G_1$ is the base point selected on first elliptic curve and $P_{b_1}$ is the $x$-coordinate of the resultant point}
				$P_{b_1} \longleftarrow Integer(P_{b_1})$	\tcp{Function to transform field to integer}
				$P_{b_2}\longleftarrow G_2 \times P_{r_2}$	\tcc{$G_2$ is the base point selected on second elliptic curve and $P_{b_2}$ is the $x$-coordinate of the resultant point}
				$P_{b_2} \longleftarrow Integer(P_{b_2})$\\	
				\KwRet{$P_{b_1}$, $P_{b_2}$}
				\label{pubKeyGen}
			\end{algorithm}

\begin{algorithm}[!ht]
 \caption{The proposed KCS-PRNG bitstream generation}
				\footnotesize{
					\KwIn{Desired length of bitstream $n$, ($574 \times r$)-bit entropy for key and IV where $r$ = $\left\lceil  \frac{n}{100000} \right\rceil$ = number of (re)seeding required for KCS-PRNG}
					\KwOut{$n$-bit cryptographically secure pseudorandom bitstream}
					Run Algorithm $\ref{ecsSel}$ to select two elliptic curves from $\mathcal{T}$ \\
					\tcp{Perform (Re)seeding to initialize the Sequence Generator}
					Run Algorithm $\ref{init}$ with input of 401-bit key and 173-bit IV to initialize the Sequence Generator\\
					Run Algorithm $\ref{seqGen}$ to generate 256-bit sequence $z_1$\\
					Transform  $z_1$ into field element $P_{r_1}$ of first elliptic curve using field converter\\
					Run Algorithm $\ref{pubKeyGen}$ with $P_{r_1}$ as input to generate the integer $P_{b_1}$\\
					Run Algorithm $\ref{seqGen}$ to generate 256-bit sequence  $z_2$\\
					Transform  $z_2$ into field element $P_{r_2}$ of second elliptic curve using field converter\\
					Run Algorithm $\ref{pubKeyGen}$ with $P_{r_2}$ as input to generate the integer $P_{b_2}$\\
					Set $countSel = 1$\\
					Set $bitCount = 1$\\
					Set $t = 1$ where $t = 1$ to $\left\lceil \frac{n}{256} \right\rceil$\\
					\For{$i \gets 1$ \textbf{to}  $\left\lceil \frac{n}{256} \right\rceil$} {
						
						\If{$countSel == t \times 256$ } {
							\tcp{Use Selector to select between the two elliptic curves}
							\eIf{$t$ is even }{
								Set $el = P_{b_2}$
							}
							{Set $el = P_{b_1}$ }
							$countSel = 0$ \\
							$t^{++}$
						}
						
						Clock Sequence Generator 256 times to generate 256-bit sequence $s$ \\
						\eIf{$n < 256$}{
							\KwRet{$X \oplus i^{th}$ position of $el$ from LSB ($i=0$) to MSB ($i=255$) where $X$ is 1-bit output from Sequence Generator and $i$ = 0 to 255} \tcp{Output of KCS-PRNG}}
						{\KwRet{$el \oplus s$} \tcp{Output of KCS-PRNG}
						}
						$i^{++}$  \\
						\If{$i == 255$ }{
							$i = 0$
						}
						$countSel^{++}$\\
						$bitCount^{++}$ \\
						\If{$bitCount == j \times 100000$ where $j = 1$ to $r$}
						{
							$n = n - (j \times 100000)$ \\
							$j^{++}$\\
							\tcp{Reseed the KCS-PRNG on every 100000 bits of output}
						Go to reseeding step in the beginning }
					}
				}
				\label{bitGen}
			\end{algorithm}

The Sequence Generator generates two sequences $z_1$ and $z_2$ of 256-bit length each, which are used by the field converter as the inputs. The field converter transforms $z_1$ and $z_2$ into integers and then transforms them into the field elements $P_{r_1}$ and $P_{r_2}$ of the two elliptic curves. These field elements or the secrets $P_{r_1}$ and $P_{r_2}$ are given as inputs to the two elliptic curve point multiplication functions as described in Algorithm $\ref{pubKeyGen}$. The secrets $P_{r_1}$ and $P_{r_2}$ are multiplied with their corresponding base points $G_1$ and $G_2$ which yields a new point on each of the elliptic curves respectively. The $x$-coordinates of the two points obtained are the two integers $P_{b_1}$ and $P_{b_2}$ after transformation from the field elements. A selector is used to switch between the outputs of the two elliptic curves point multiplication functions to double the size of key space offered by the proposed KCS-PRNG.

\medskip	
Algorithm $\ref{bitGen}$ describes the cryptographically secure pseudorandom bitstream generation scheme of the proposed KCS-PRNG. Initially, two elliptic curves with hard ECDLP are selected from $\mathcal{T}$. The Sequence Generator is then initialized with 401-bit key and 173-bit IV as discussed in Algorithm $\ref{init}$. The Sequence Generator is used to generate 256-bit sequence $z_1$ by clocking 256 times. Further, $z_1$ is converted into the field element of the first elliptic curve and considered as the secret $P_{r_1}$.~The integer $P_{b_1}$ is generated by using elliptic curve point multiplication function taking the secret $P_{r_1}$ as input. Similarly, the integer $P_{b_2}$ is also generated from the second elliptic curve point multiplication function. The Sequence Generator continuously generates $n$-bit length sequences as bounded by $\left\lceil \frac{n}{256} \right\rceil$ times loop. The proposed KCS-PRNG uses a selector iteratively select among $P_{b_1}$ and $P_{b_2}$. The Sequence Generator is then clocked 256 times to generate 256-bit sequence $s$. The integers $P_{b_1}$ or $P_{b_2}$ is masked with $s$ to produce 256-bit output by the KCS-PRNG. If $n < 256$, then 1-bit output of the Sequence Generator is masked with 1-bit of $P_{b_1}$ or $P_{b_2}$ (as decided by the selector) traversing from its Least Significant Bit (LSB) to MSB and result is returned. Once MSB of the $P_{b_1}$ or $P_{b_2}$ is used, the masking of the output of the Sequence Generator starts from the LSB of the $P_{b_1}$ or $P_{b_2}$ once again in rotating fashion. The KCS-PRNG is reseeded on every 100000 bit of output to maintain backward secrecy as shown in Algorithm $\ref{bitGen}$.
			
\subsection{Assumptions}

Following assumptions are made in the proposed design of KCS-PRNG:
			\begin{itemize}
 \itemsep=0.85pt
				\item{KCS-PRNG always maintains 574-bit initial entropy.}
				\item{KCS-PRNG expects high per-bit entropy $\approxeq$ 1 for initialization. The generation details of entropy used in KCS-PRNG is outside the scope of this work.}
				\item{The Key and IV are parts of the seed and hence, they are immediately shredded after use and is non-recoverable.}
				\item{The (Re)keying and (Re)IVing are done using different TRNGs or entropy harvesters using various different physical noise sources.}
				\item{Elliptic curves used in KCS-PRNG are randomly generated, cryptographically safe and trustworthy.}
				\item{Look-up Table $\mathcal{T}$ has authorized access only.}
			\end{itemize}
			

\section{Security analysis of the proposed KCS-PRNG} \label{sec5}
		
\subsection{Linear complexity analysis}

Let linear complexities of the Sequence Generators $SG_1, SG_2$ and $SG_3$ be $LC_1, LC_2$ and $LC_3$ respectively and following equation ($\ref{nonLinearFn}$), are given by
		\begin{equation} \label{linSGs}
			\begin{array}{lll}
				LC_1 = L_1L_2 + L_2L_3 + L_1L_3 = 3119 \\
				LC_2 = L_4L_5 + L_5L_6 + L_4L_6 = 5711 \\
				LC_3 = L_7L_8 + L_8L_9 + L_7L_9 = 9959
			\end{array}
		\end{equation}
		where $L_1, L_2, \cdots, L_9 $ are the lengths of the LFSRs.
\eject

		Moreover, while $SG_1$ is clocked regularly, $SG_2$ and $SG_3$ are connected in alternating step configuration. Thus, following equation ($\ref{linComplexity}$), the overall linear complexity ($LC$) of the scheme is given by
		\begin{equation} \label{linSG}
			\begin{array}{l}
				(5711 + 9959)^{2 \times 3119-1} < LC(x) \leq (5711 + 9959)^{2 \times 3119}	\\
				\implies	15670^{6237} < LC(x) \leq 15670^{6238}
			\end{array}
		\end{equation}
		It is imperative to note that the Sequence Generator of the proposed KCS-PRNG exhibits exponentially large linear complexity as demonstrated in equation ($\ref{linSG}$) and therefore, the proposed generator is resistant to the Berlekamp-Massey attack \cite{menezes2018handbook}.

\begin{table}[!b]
\vspace*{-2mm}
			\centering
			\scriptsize{
				\caption{Correlation test of the proposed KCS-PRNG.} \label{corrTest}\vspace*{-3mm}
				\begin{tabular}{|l|l|}
						\hline
						\begin{tabular}[c]{@{}l@{}} sstring-AutoCor test \end{tabular} & \begin{tabular}[c]{@{}l@{}}$N=1, n=1048513, r=0, s=32, d=1$ \end{tabular} \\ \hline \hline
						Normal statistic & 0.41 \\  \hline
						p-value of test &	0.34 \\  \hline
						Number of bits used & 1048544 \\  \hline
						Result &	Passed the test \\  \hline
						sstring-AutoCor test & $N=1, n=1048514, r=0, s=32, d=2$ \\  \hline
						Normal statistic & 0.80 \\  \hline
						p-value of test &	0.21 \\  \hline
						Number of bits used & 1048544 \\  \hline
						Result &	Passed the test \\ \hline \hline
						sstring-HammingCorr test & $N=1, n=32768, r=0, s=32, L=32$ \\  \hline  \hline
						Normal statistic & -0.56 \\  \hline
						p-value of test &	0.71 \\  \hline
						Number of bits used & 1048576 \\  \hline
						Result &	Passed the test \\ \hline  \hline
						sstring-HammingCorr test & $N=1, n=16384, r=0, s=32, L=64$ \\ \hline  \hline
						Normal statistic & 0.45 \\  \hline
						p-value of test &	0.33 \\  \hline
						Number of bits used & 1048576 \\  \hline
						Result &	Passed the test \\ \hline \hline
						sstring-HammingCorr test & $N=1, n=8192, r=0, s=32, L=128$ \\ \hline \hline
						Normal statistic & 1.57 \\  \hline
						p-value of test &	0.06 \\  \hline
						Number of bits used & 1048576 \\  \hline
						Result &	Passed the test \\  \hline
					\end{tabular}
				}
			\end{table}

\subsection{Correlations test}

We conducted two correlation tests of random bitstreams generated by the proposed KCS-PRNG to verify non-correlation in the bitstream. The first test conducted was Serial or Autocorrelation test ($sstring-AutoCor$ test) which measures the correlation between the bits with the lag $d$ \cite{l2007testu01}. In this test, a $n$-bit string is generated by the KCS-PRNG at the first level and the test statistic is computed such that it has the binomial distribution with the parameters being approximately standard normal for large $n-d$. The restriction imposed were $r+s \leq 32$ and $1 \leq d \leq \lfloor \frac{n}{2} \rfloor$ where $r$ be the number of MSBs which are eliminated from the output before applying the test, $s$ be the MSBs chosen from each generated random number and $N$ be second-level number of replications \cite{l2007testu01,l2013testu01}. The second test conducted was the Hamming Correlation test ($sstring-HammingCorr$) \cite{walker2008ent} was used to measure bitwise correlation in the random bitstream file of 1GB size generated by the proposed KCS-PRNG which was estimated to be 0.000034. The obtained correlation is very close to the ideal correlation value of 0.0 and thus, concludes that the proposed design of the KCS-PRNG has no correlation issues and their results are shown in Table $\ref{corrTest}$.

\subsection{Period analysis (validation of requirement R$\ref{R1}$)}

The Sequence Generator used in the KCS-PRNG comprises of nine LFSRs whose lengths $L_1, L_2, \cdots, L_9$ are coprime to each other. Hence, the period ($P$) of the Sequence Generator is given by
			\begin{equation}	\label{period1}
				P = \prod\limits_{i=1}^9 (2^{L_i} - 1)
			\end{equation}
			As the output of the Sequence Generator is masked with the integer obtained from the $x$-coordinate of the public key of one of the two elliptic curves, therefore, the period $\mathcal{P}$ of the proposed KCS-PRNG is given by
			\begin{equation}	\label{periodKCSPRNG}
				\mathcal{P} = \left\{
				\begin{array}{ll}
					N_1 \times (\prod\limits_{i=1}^9 (2^{L_i} - 1)) & \quad if (n \leq 256) \\
					(N_1 + N_2) \times (\prod\limits_{i=1}^9 (2^{L_i} - 1)) &  \quad if (n > 256)
				\end{array}
				\right.
			\end{equation}
			where $n$ be the number of output bits and $N_1, N_2$ are the order of the two elliptic curves and let $N_1 < N_2$.

\medskip			
			It is prudent from equation ($\ref{periodKCSPRNG}$) that the period $\mathcal{P}$ of the proposed KCS-PRNG approximately lies in the range $[N_1 \times 2^{401}, (N_1 + N_2) \times 2^{401}]$ per boot which enables the proposed KCS-PRNG to generate very large bitstream without compromising the statistical properties of randomness.
			
			\subsection{Key space analysis}
			It is evident from equation ($\ref{period1}$) that the Sequence Generator in KCS-PRNG has a period of $2^{401}$ and thus, provides $2^{401}$ key space in case the generator gets seeded once and no reseeding happens. Moreover, the KCS-PRNG also uses two elliptic curves which provides $2^{128}$ and $2^{256}$ key space for $n \leq 256$ and $n > 256$ bits of output respectively to impose a successful Pollard's rho attack to solve the ECDLP. Hence the key space offered by the proposed KCS-PRNG is given by
			\begin{equation}	\label{keySpace}
				\mathcal{K} = \left\{
				\begin{array}{ll}
					(2^{401} \times 2^{128})^r = 2^{529r} & \quad if (n \leq 256) \\
					(2^{401} \times 2^{256})^r = 2^{657r} &  \quad if (n > 256)
				\end{array}
				\right.
			\end{equation}
			where $r$ be the number of (re)seeding the KCS-PRNG and $n$ be the number of output bits of the proposed KCS-PRNG.
	
\medskip		
It is imperative to note that the key space offered by the proposed KCS-PRNG depends on the number of times the KCS-PRNG (re)seeds itself in single boot and therefore, exhibits virtually infinite key space in the range $\mathcal{K} \in [2^{529}, \infty)$ which is quite higher than the safe key space threshold of $2^{128}$ as recommended by \cite{alhadawi2019designing,ii2010ecrypt}. Therefore, the proposed KCS-PRNG comfortably resists brute force\linebreak attacks.
			
\subsection{Cross layer attack on kernel PRNG}

A practical attack \cite{klein2021cross} using the weakness in the Linux Kernel PRNG is discovered that allowed the hackers to mount the cross-layer attacks against the Linux kernel to retrieve the internal states of the PRNG.~The internal states of the kernel PRNG were compromised due to the linearity, same set of instances being used by the applications of the kernel PRNG and partially re-seeding issues respectively.~The attackers were able to extract data from one PRNG consumer (network protocols like IPv4/IPv6, UDP etc.) in one Open Systems Interconnection (OSI) layer and used them to exploit another PRNG consumer in difference OSI layer.~This weakness in the PRNG also allowed hackers to identify and track both the Linux and the Android devices.~The compromised kernel was then used to downgrade E-mail security, hijack E-mails, hijack Hyper Text Transport Protocol (HTTP) traffic, circumvent E-mail anti-spam and blacklisting mechanisms, mount a local Denial of Service (DoS) attack (blackhole hosts), poison reverse Domain Name Server (DNS) resolutions and attack the machine’s Network Time Protocol (NTP) client which is responsible for the machine’s \linebreak clock.

\medskip			
			It is imperative to note that the compromised internal states of the kernel PRNG enabled the attackers to predict entire random sequences generated by it.~However, the proposed KCS-PRNG does not allow such leakage of its internal states due to its unique design that leverages very high degree of non-linearity (as given in equation ($\ref{linSG}$)) of the generator and generates non-reproducible random bit sequences to provide entirely unrelated pseudorandom sequences for each user applications.
			
\section{Experimental validation of the proposed KCS-PRNG} \label{sec6}

\subsection{Experimental validation of requirement $R\ref{R1}$}

\begin{itemize}
	\item[i.]	NIST statistical test results	 \newline
				NIST test suite consists of 15 statistical tests to certify statistical strength of randomness of the RNG. An output bitstream of 1GB file size is generated by the proposed KCS-PRNG and subjected to the NIST tests using NIST statistical test suite SP 800-22 version 2.1.2 \cite{bassham2010sp}. The input block size was set to be 1000000 bits and 1000 bitstreams. The significance level $\alpha$ was selected as 99\% to conduct the test. The proposed KCS-PRNG passed all the NIST statistical tests and the details of test results obtained are depicted in Table $\ref{nist}$.
		
\medskip		
				The p-value measures randomness and supposed to be greater than 0.01 i.e., the confidence level to conclude that the sequence is uniformly distributed whereas the proportion i.e., the minimum pass rate for the test should fall in the range [0.98056, 0.99943] having the confidence interval $\alpha$=0.01 and 1000 bitstreams \cite{anandakumar2019fpga}. As indicated in Table $\ref{nist}$, the proposed KCS-PRNG not only qualifies the pass rate threshold of 0.98056 but also reports better pass rate of 0.9896 as compared to the pass rates of 0.987 and 0.9887 reported by the TRNG \cite{anandakumar2019fpga} and the PRNG \cite{alhadawi2019designing} respectively.

				\begin{table}[!ht]
					\centering
					\small{
			        \caption{NIST test results of the proposed KCS-PRNG output bitstreams of 1GB file size
                     with the input of 1000000-bit block size and 1000 bitstreams.}
						\begin{tabular}{|l|l|l|l|}
							\hline
							Statistical Test & $P-value$  & Proportion & Result \\ \hline \hline
							Frequency	&	0.737915       & 0.991	& Pass  \\ \hline
							Block Frequency & 0.591409 & 0.988 & Pass  \\	 \hline
							CumulativeSums*	 & 0.680755 & 0.993	& Pass  \\ \hline
							Runs	&	0.281232           & 0.992	& Pass  \\  \hline
							Longest Run	&	0.526105     & 0.996	& Pass  \\  \hline
							Rank	&	0.036113           & 0.996	& Pass  \\  \hline
							FFT	&	0.103138             & 0.990	& Pass  \\  \hline
							NonOverlappingTemplate* 	 & 0.794391  & 0.990 & Pass  \\  \hline
							Overlapping	&	0.779188     & 0.987	& Pass  \\  \hline
							Universal	&	0.773405       & 0.991	& Pass  \\  \hline
							Approx Entropy &	0.653773 & 0.989	& Pass  \\  \hline
							RandomExcursions* &	0.489508    & 0.983	& Pass  \\  \hline
							RandomExcursionsVariant*	&	0.163362         & 0.985	& Pass  \\  \hline
							Serial*	&	0.680755        & 0.988	& Pass  \\  \hline
							Linear Complexity	&	0.682823              & 0.985	& Pass  \\ \hline
						\end{tabular}
						\label{nist} \newline\newline
						*Only the result of first test instance is indicated here from the original results due to limitation of space
					}
				\end{table}
				
				\item[ii.]	Diehard test results \cite{marsaglia1998diehard}	\newline
				Diehard version 3.31.1 tests conduct a series of statistical tests and determine the p-values of the output bitstreams. The p-values indicate deviation of bit prediction from ideally expected probability of half. The expected p-value of a test should be in the range [0.025, 0.975] \cite{bhattacharjee2018search}. The proposed KCS-PRNG passed all the diehard tests as shown in Table $\ref{diehard}$.

				\begin{table}[!ht]
					\centering
					\small{
						\caption{Diehard test results of the proposed KCS-PRNG output bitstreams of 1GB file size.}\vspace*{-2mm}
				\scalebox{0.9}{
             		\begin{tabular}{|l |c |c |c |c |c|}
							\hline
							test-name   & ntup & tsamples  & psamples &  p-value  & Assessment\\ \hline \hline
							diehard-birthdays &   0 &       100 &     100 & 0.27561288 &  Passed  \\  \hline
							diehard-operm5 &   0 &   1000000 &     100 & 0.13184067 &  Passed  \\  \hline
							diehard-rank-32x32 &   0 &     40000 &     100 & 0.44295780 &  Passed  \\  \hline
							diehard-rank-6x8 &   0 &    100000 &     100 & 0.88076181 &  Passed  \\  \hline
							diehard-bitstream &   0 &   2097152 &    100 & 0.42947798 &  Passed  \\  \hline
							diehard-opso &   0 &   2097152 &     100 & 0.12604767 &  Passed  \\  \hline
							diehard-oqso &   0 &   2097152 &     100 & 0.94641900 &  Passed  \\  \hline
							diehard-dna &   0 &   2097152 &   100 & 0.24390543 &  Passed  \\  \hline
							diehard-count-1s-str &   0 &    256000 &     100 & 0.62287409 &  Passed  \\  \hline
							diehard-count-1s-byt &  0 &    256000 &     100 & 0.91047395 &  Passed  \\  \hline
							diehard-parking-lot &   0 &     12000 &     100 & 0.79390338 &  Passed  \\  \hline
							diehard-2dsphere &   2 &      8000 &     100 & 0.17731451 &  Passed  \\  \hline
							diehard-3dsphere &   3 &      4000 &     100 & 0.45129204 &  Passed  \\ \hline
							diehard-squeeze &   0 &    100000 &     100 & 0.53561994 &  Passed  \\  \hline
							diehard-sums &   0 &       100 &    100 & 0.94209561 &  Passed  \\  \hline
							diehard-runs* &  0 &    100000 &    100 & 0.14811353 &  Passed  \\  \hline
							diehard-craps* &   0 &    200000 &    100 & 0.92115680 &  Passed  \\  \hline
							marsaglia-tsang-gcd* &   0 &  10000000 &     100 & 0.53120802 &  Passed  \\  \hline
							sts-monobit &    1 &    100000 &     100 & 0.64501072 &  Passed  \\  \hline
							sts-runs &   2 &    100000 &     100 & 0.94961272 &  Passed  \\  \hline
							sts-serial* &   1 &  100000 &     100 & 0.62077367 &  Passed  \\  \hline
							rgb-bitdist* &   1 &    100000 &     100 & 0.95378266 &  Passed \\  \hline
							rgb-minimum-distance* &   2 &    10000 &    1000 & 0.87517368 &  Passed  \\  \hline
							rgb-permutations* &   2 &    100000 &     100 & 0.75286377 &  Passed  \\  \hline
							rgb-lagged-sum* &   0 &   1000000 &     100 & 0.00308570 &  Passed  \\  \hline
							rgb-kstest-test &   0 &     10000 &    1000 & 0.03414230 &  Passed  \\  \hline
							dab-bytedistrib &   0 &  51200000 &      1 & 0.17158919 &  Passed  \\  \hline
							dab-dct & 256 &     50000 &       1 & 0.07312246 &  Passed  \\  \hline
							dab-filltree* &  32 & 15000000 &       1 & 0.61801753 &  Passed  \\  \hline
							dab-filltree2* &   0 &   5000000 &       1 & 0.69361846 &  Passed  \\  \hline
							dab-monobit2 &  12 &  65000000 &       1 & 0.42742922 &  Passed \\ \hline
						\end{tabular} }
						\label{diehard} \newline\newline
						*Only the result of first test instance is indicated here from the original results due to limitation of space
					}
				\end{table}

				\item[iii.]	TestU01 test results \cite{l2007testu01}	\newline
				TestU01 is believed to impose the toughest tests to evaluate the statistical quality of random bitstreams \cite{alhadawi2019designing}. The binary bitstream of 1GB file size generated by the proposed KCS-PRNG is subjected to the Rabbit and Alphabit test batteries of TestU01. The Rabbit and the Alphabit, by default, selected 1048576 bits ($2^{20}$ bits) for SmallCrush (a fast statistical test battery) evaluation and applied 38 and 17 statistical tests respectively to the proposed KCS-PRNG output bitstream. The output bitstreams of KCS-PRNG are found to have p-values within the acceptable range of [0.001, 0.999] \cite{bhattacharjee2018search} which proved that the proposed KCS-PRNG exhibits long period, good structure and non-linearity.
			\end{itemize}
			
\subsection{Validation of requirements R$\ref{R2}$ and R$\ref{R3}$}

\begin{itemize}
				\item[i.]	Next bit test \newline
				This test states that if a sequence of $m$-bits is generated by a generator, there should not be any feasible method which can predict the ($m+1$)th bit with the probability significantly higher than half \cite{lavasani2009practical,Rose2011crypto}. This test is associated with predictability of the successive bits generated by the KCS-PRNG.
				
				Since the KCS-PRNG is reseeded with fresh additional entropy of 574 bits (401 bits of key and 173 bits of IV), therefore, it maintains backward security \cite{ferguson2011cryptography}.
				
				\item[ii.]	Test for state compromise extension attacks	\newline
				This test states that if some state of a generator is leaked at a given time to an attacker, it would not be possible to recover unknown PRNG outputs from that known state \cite{kelsey1998cryptanalytic}. Fundamentally, the state compromise extension imposes two kinds of attack: first, a backtracking attack to learn previous outputs of the generator knowing some internal state of the generator at a particular time and second, the permanent compromise attack which enables all the future and past states of the generator vulnerable with the knowledge of some state at a given time \cite{kelsey1998cryptanalytic}.
				
				Since the proposed KCS-PRNG is forward secure and provably secure due to underlying ECDLP intractability, therefore, it is resistant to the backtracking attack. Furthermore, as discussed in the next bit test, the proposed KCS-PRNG is (re)seeded on every 100000 bits of output generation, therefore, it exhibits backward secrecy and thus, resists the permanent compromise attack as well.
				
				\item[iii.] Entropy Estimation (Experimental Validation of Requirement R$\ref{R2}$, R$\ref{R3}$)
				Entropy is the measurement of unpredictability or uncertainty. For an ideal TRNG, the expected entropy is 1 per bit which means that each bit i.e., `0' or `1' have equal proportion 0.5 in the file containing random bitstream \cite{anandakumar2019fpga}. The proposed KCS-PRNG is subjected to ENT tool \cite{walker2008ent} for estimation of the entropy of the KCS-PRNG generated 1GB file of random bitstream. The observed value of the entropy of output bitstream generated by the proposed KCS-PRNG is found to be 0.99999975 per bit which asserts that the design of KCS-PRNG maintains nearly an ideal unpredictability.
			\end{itemize}

\subsection{Experimental validation of requirement R$\ref{R4}$}
			
\subsubsection{Non-reproducibility test}

The non-reproducibility test is conducted to validate if the RNG requirement $R\ref{R4}$ is met by the proposed KCS-PRNG. This test is conducted by running the generator twice with exactly the same input and verifying if the output sequences are completely unrelated. Authors \cite{anandakumar2019fpga} have referred the non-reproducibility test as the restart test and they validated the first 20 bit output sequences of the generator six times under identical start conditions. Table $\ref{nonReprTest}$ shows that the proposed KCS-PRNG has passed the non-reproducibility test six times by producing six completely unrelated 32 bits using the same inputs to the proposed generator.

				Moreover, the KCS-PRNG uses two different elliptic curves on each boot and therefore, the output bitstream would be entirely unrelated even generated under identical start conditions. Hence, it is inferred that the proposed KCS-PRNG generates non-reproducible pseudorandom bitstreams, provided it maintains minimum number of un-used elliptic curves (i.e., $t+1$ where $t \ge 1$ is the number of (re)boots made by the KCS-PRNG such that the generator gets at least two un-used elliptic curve on each (re)boot) in the look-up table consisting of elliptic curves.

\clearpage

	\begin{table}[!ht]
   \vspace*{-2mm}
				\centering
				\footnotesize{
	\caption{Non-reproducibility test of the proposed KCS-PRNG under identical start conditions.}\vspace*{-2mm}
	\scalebox{0.88}{
          \begin{tabular}{|l|l|}
			\hline
				\begin{tabular}[c]{@{}l@{}}Key Input (401-bit entropy)\end{tabular} & \begin{tabular}[c]{@{}l@{}}1905119BCDC809077DB45D \\  1B3921DB5C06D11 C56C7FE \\  B4F8EE935A2FB16B055281816\\  DFC551AC73C3BBF76EE26B13 \\  0B8F5E68 \end{tabular}\\ \hline
							\begin{tabular}[c]{@{}l@{}} IV Input (173-bit entropy)\end{tabular} & \begin{tabular}[c]{@{}l@{}}190B6B491CDD9E97E6AB \\  26552990F5481183DEF9AE55\end{tabular}\\ \hline \hline
							Check & First run of KCS-PRNG \\  \hline
							32-bit Output &	01010100111011111110001110100100 \\  \hline
							Check & Second run of KCS-PRNG \\   \hline
							32-bit Output &	00010010000100001111001111111110 \\  \hline
							Check & Third run of KCS-PRNG \\  \hline
							32-bit Output &	11000101110001101011100101111101 \\  \hline
							Check & Fourth run of KCS-PRNG \\  \hline
							32-bit Output &	01101010010110101011000010110101  \\  \hline
							Check & Fifth run of KCS-PRNG \\ \hline
							32-bit Output &	10110001000111011001101100011011 \\  \hline
							Check & Sixth run of KCS-PRNG \\  \hline
							32-bit Output &	01001100110010111100010011100110 \\ \hline
						\end{tabular} }
						\label{nonReprTest}\vspace*{6mm}
					}

					\centering
					\scriptsize{
						\caption{Details of first elliptic curve with verification details \cite{dj2015safe} used in the proposed KCS-PRNG}\vspace*{-2mm}
						\scalebox{0.98}{	\begin{tabular}{|l |l|}
							\hline
							Elliptic curve parameter/Validation & Value \\ \hline \hline
							Equation Model & Short Weierstrass \\  \hline
							Prime field $p$ &	 0xEEAA0DB0A46CE48AFCD288C714939E4063E1D801C55D1118202C76798B62B483 \\ \hline
							Coefficient $a$ &	 0x33866AAA5914BC27D9ED986D7AF431BD8FC217D8E07D5BA5E44C1A4A355C7DD4\!\! \\ \hline
							Coefficient $b$ &	 0xCAA0537DF123F85EC185A991B7200396B996C7921E6A7E07F08ED2A4801B0CA2 \\  \hline
							Co-factor $h$ &	0x1  \\  \hline
							Base Point $G_x$ & 0x3FBE1FF3CC8A893B2B018CC7D3D61961233F87F66FCB257D21805D1327426DE9 \\  \hline
							Base Point $G_y$ & 0xC5B219E84B008A4CB36CDF05B44E95354913756FCD92251F90BFB0A4F4D84AD8 \\  \hline
							$Rho$ &	 \begin{tabular}[c]{@{}l@{}}127.8  \tcp{Key space of $2^{127.8}$ for Pollard's Rho attack on ECDLP}\end{tabular} \\  \hline
							$Twist-rho$ &  \begin{tabular}[c]{@{}l@{}}	 127.8 \end{tabular}  \\  \hline
							$Joint-rho$ &  \begin{tabular}[c]{@{}l@{}}	 127.8 \end{tabular} \\  \hline
							$verify-isElliptic$ &  \begin{tabular}[c]{@{}l@{}}	 True \tcp{Ensuring elliptic curve} \end{tabular} \\  \hline
							$verify-P_{r_1}isOnCurve$ &  \begin{tabular}[c]{@{}l@{}}	True \tcp{Ensuring private key as an elliptic curve group element} \end{tabular} \\  \hline
							$verify-P_{b_1}isOnCurve$ &	 \begin{tabular}[c]{@{}l@{}}True \tcp{Ensuring public key as an elliptic curve group element}\end{tabular}  \\  \hline
							$verify-safeField$ &  \begin{tabular}[c]{@{}l@{}}True	 \tcp{Elliptic curve defined over a suitable prime field} \end{tabular} \\  \hline
							$verify-safeEquation$ &  \begin{tabular}[c]{@{}l@{}}	 True \tcp{Ensuring short Weierstrass equation} \end{tabular} \\  \hline
							$verify-safeBase$ &  \begin{tabular}[c]{@{}l@{}}	 True \tcp{Ensuring base point with prime order}\end{tabular}  \\  \hline
							$verify-safeRho$ &  \begin{tabular}[c]{@{}l@{}}	True \tcp{Ensuring ECDLP security} \end{tabular} \\
							$verify-safeTransfer$ &  \begin{tabular}[c]{@{}l@{}}	True \tcp{Ensuring ECDLP security from MOV attacks} \end{tabular} \\  \hline
							\begin{tabular}[c]{@{}l@{}}$verify-safeDiscriminant$ \end{tabular}&  \begin{tabular}[c]{@{}l@{}}	True \tcp{Ensuring cubic curve} \end{tabular} \\ \hline
							$verify-safeRigid$ &  \begin{tabular}[c]{@{}l@{}}	True \tcp{Ensuring elliptic curve is generated using explained procedure}\end{tabular}  \\  \hline
							$verify-safeTwist$ &  \begin{tabular}[c]{@{}l@{}}	True \tcp{Ensuring twist of the elliptic curve is safe} \end{tabular} \\  \hline
							$verify-safeCurve$ &  \begin{tabular}[c]{@{}l@{}}	True \tcp{if and only if all the other validations return `True'} \end{tabular} \\ \hline
						\end{tabular} }
						\label{EC1Details}
					}
				\end{table}

			\begin{table}[!ht]
					\centering
					\scriptsize{
						\caption{Details of second elliptic curve with verification details \cite{dj2015safe} used in the proposed KCS-PRNG}\vspace*{-2mm}
					\scalebox{0.98}{	\begin{tabular}{|l|l|}
							\hline
							\begin{tabular}[c]{@{}l@{}} Elliptic curve parameter/\\Validation \end{tabular} & \begin{tabular}[c]{@{}l@{}}Value  \end{tabular}\\ \hline \hline
							\begin{tabular}[c]{@{}l@{}} Equation Model \end{tabular} & \begin{tabular}[c]{@{}l@{}}Short Weierstrass \end{tabular} \\  \hline
							\begin{tabular}[c]{@{}l@{}} Prime field $p$  \end{tabular} &	0xF2A284E729748EA8BE82173F13412FC257C42095408D706528F5D8964BF2E237  \\  \hline
							\begin{tabular}[c]{@{}l@{}} Coefficient $a$ \end{tabular}  &	0xB29C202E105FE4C7EE5DECAF48258BFAB2E890AF5D96DE4553D82C3EC5D03C06 \\  \hline
							\begin{tabular}[c]{@{}l@{}} Coefficient $b$  \end{tabular} &	0xC36BBDD9EE50EF046EA1D4DA85300673531B323B013043F9DC97B2FDD6A807B4 \\  \hline
							\begin{tabular}[c]{@{}l@{}} Co-factor $h$ \end{tabular}  &	 0x1 \\  \hline
							\begin{tabular}[c]{@{}l@{}} Base Point $G_x$  \end{tabular} & 0x1216C78C1FB8707C6B7B2496226B6F13CE25347DD9283A36FA354D09E2CDF4C3 \\  \hline
							\begin{tabular}[c]{@{}l@{}} Base Point $G_y$  \end{tabular} & 0xA0AC0431A50C5DA5D25DCA1026946A2AADA19756ED326DA85A203B4A0B2BE342 \\  \hline
							$Rho$ &	 \begin{tabular}[c]{@{}l@{}}127.8  \tcp{Key space of $2^{127.8}$ for Pollard's Rho attack on ECDLP} \end{tabular}\\ \hline
							$Twist-rho$ & \begin{tabular}[c]{@{}l@{}}	 127.8  \end{tabular}\\  \hline
							$Joint-rho$ & \begin{tabular}[c]{@{}l@{}}	 127.8 \end{tabular} \\  \hline
							$verify-isElliptic$ & \begin{tabular}[c]{@{}l@{}}	 True \tcp{Ensuring elliptic curve}\end{tabular} \\  \hline
							$verify-P_{r_1}isOnCurve$ & \begin{tabular}[c]{@{}l@{}}	True \tcp{Ensuring private key as an elliptic curve group element}\end{tabular} \\ \hline
							$verify-P_{b_1}isOnCurve$ & \begin{tabular}[c]{@{}l@{}}	True \tcp{Ensuring public key as an elliptic curve group element} \end{tabular}\\  \hline
							$verify-safeField$ &  \begin{tabular}[c]{@{}l@{}}True	 \tcp{Elliptic curve defined over a suitable prime field} \end{tabular}\\  \hline
							$verify-safeEquation$ & \begin{tabular}[c]{@{}l@{}}	 True \tcp{Ensuring short Weierstrass equation} \end{tabular}\\  \hline
							$verify-safeBase$ &	\begin{tabular}[c]{@{}l@{}} True \tcp{Ensuring base point with prime order} \end{tabular}\\  \hline
							$verify-safeRho$ &	\begin{tabular}[c]{@{}l@{}}True \tcp{Ensuring ECDLP security} \end{tabular}\\  \hline
							$verify-safeTransfer$ & \begin{tabular}[c]{@{}l@{}}	True \tcp{Ensuring ECDLP security from MOV attacks}\end{tabular} \\  \hline
							$verify-safeDiscriminant$ & \begin{tabular}[c]{@{}l@{}}	True \tcp{Ensuring cubic curve} \end{tabular}\\  \hline
							$verify-safeRigid$ & \begin{tabular}[c]{@{}l@{}}	True \tcp{Ensuring elliptic curve is generated using explained procedure} \end{tabular}\\  \hline
							$verify-safeTwist$ &	 \begin{tabular}[c]{@{}l@{}} True \tcp{Ensuring twist of the elliptic curve is safe} \end{tabular}\\ \hline
							\begin{tabular}[c]{@{}l@{}}$verify-safeCurve$ \end{tabular}& \begin{tabular}[c]{@{}l@{}} True \tcp{if and only if all the other validations return `True'} \end{tabular}\\ \hline
						\end{tabular} }
						\label{EC2Details}
					}\vspace*{2mm}
				\end{table}

\section{Details of two elliptic curves used in the proposed KCS-PRNG}	\label{sec7}

Elliptic curves over 256-bit prime fields whose ECDLPs are found to be hard and method of computation is transparent and trustworthy, are selected for use in the proposed KCS-PRNG. The elliptic curves are generated randomly over the 256-bit prime field size in order to build the trust as indicated in \cite{bernstein2015manipulate,shumow2007possibility,hales2013nsa,bernstein2013security}. The generation mechanism of cryptographically safe elliptic curves is referred from \cite{dj2015safe,konstantinou2010efficient,menezes1993reducing,cheng2008hard,bos2016selecting,smart1999discrete,koblitz2004guide} and followed with the procedure suggested in \cite{abhishek2021computation} to achieve trusted security. The proposed KCS-PRNG uses two elliptic curves which are generated randomly and verified for their cryptographic security as per the recommendations given in \cite{dj2015safe}. The verification details against the criteria as suggested in \cite{dj2015safe} of the two elliptic curves selected for experimentation purposes in this work are summarized in Table $\ref{EC1Details}$ and Table $\ref{EC2Details}$ respectively. A look-up table $\mathcal{T}$ used in the proposed KCS-PRNG is created with 256 such elliptic curves initially as discussed in Section $\ref{architecture}$.

\section{Results}\label{sec8}

The security analysis carried out in Section $\ref{sec5}$ has proved that the proposed KCS-PRNG exhibits: higher security property (from RNG requirements $R\ref{R1}$ to $R\ref{R4}$), provably secure, very high per bit entropy rate, minimal bitwise correlation, highly nonlinear with linear complexity $LC(x)$ bounded as $15670^{6237} < LC(x) \leq 15670^{6238}$, very large period in the range $[N_1 \times 2^{401}, (N_1 + N_2) \times 2^{401}]$ per boot where $N_1 < N_2$ being the order of two elliptic curves used, huge key space in the range $[2^{529}, \infty)$ and impressive throughput of 2.5 Megabits per second as discussed in Section $\ref{sec9}$ to generate uninterrupted cryptographically secure bitstreams.

\medskip				
The proposed KCS-PRNG passed all the tests of NIST, Diehard and TestU01 test suites along with other tests to validate statistical qualities of randomness, cryptographic security and non-reproducibility as discussed in Section $\ref{sec6}$. The NIST test also proved that the proposed KCS-PRNG exhibits impressive and the highest proportion i.e., the pass rate of 0.9896 as compared to the existing PRNG \cite{alhadawi2019designing} with 0.9887 and TRNG \cite{anandakumar2019fpga} with 0.987 proportion values. The KCS-PRNG demonstrated to exhibit nearly an ideal 0.99999975 per bit entropy and minimal serial correlation of 0.000034 in its generated bitstream.

\section{Performance analysis of the proposed KCS-PRNG} \label{sec9}\vspace{0.5mm}
				
The proposed KCS-PRNG was run on Intel$\textsuperscript{\textregistered}$ Core$\textsuperscript{TM}$ i7-7700 CPU @ 3.60GHz processor. The source code of the KCS-PRNG is developed in C++ and extensively used CryptoPP version 8.2.1 library. The KCS-PRNG software program was run on Ubuntu version 16.04.1 with kernel version 4.15.0-96-generic. The KCS-PRNG program was (re)seeded on every 100000 bits output in generation of 1GB file of pseudorandom bitstream. It gave an impressive throughput of 2.5 Mbps in software which asserts its high throughput-oriented design. The proposed KCS-PRNG for kernel applications offers a better security by meeting all the RNG requirements from $R\ref{R1}$ to $R\ref{R4}$ as compared to the existing PRNG \cite{alhadawi2019designing} and kernel CSPRNGs like/dev/random \cite{viega2003practical,dodis2013security}, Yarrow \cite{kelsey1999yarrow}, and \linebreak Fortuna \cite{ferguson2011cryptography}.

\section{Comparison of proposed KCS-PRNG with recent CSPRNGs for kernel applications} \label{sec10}\vspace{0.5mm}

The proposed KCS-PRNG is designed to meet all the requirements of a RNG as discussed in Section $\ref{subsec4}$. The features of the proposed KCS-PRNG are compared with the popular CSPRNGs used by the current operating system kernels and a recently well acknowledged TRNG \cite{anandakumar2019fpga} in Table $\ref{compCSPRNGs1}$. The reason behind the comparison of KCS-PRNG with TRNG is that, it meets the RNG requirement $R\ref{R4}$ which a TRNG only meets. Table $\ref{compCSPRNGs1}$ also consolidates interesting comparison results of KCS-PRNG with an existing TRNG based on Oscillator-Rings \cite{anandakumar2019fpga}.
				
\medskip
The KCS-PRNG is compared with popular kernel CSPRNGs namely /dev/(u)random used by Linux and Android kernels, Yarrow used by MacOS/iOS/FreeBSD kernel and Fortuna used by Windows kernel respectively on the basis of various criteria related to cryptographic security, randomness tests and throughput to conclude their suitability for strategic applicatons such as kernel applications.

				\begingroup	
				\setlength{\tabcolsep}{7.5pt} 
				\renewcommand{\arraystretch}{1.0} 
				\begin{landscape}	\hbox{}\hspace*{80mm}	
					\begin{center}
					\begin{table}[h!]
							\centering
							\footnotesize{
								\caption{Comparison of the proposed KCS-PRNG with recent Kernel CSPRNGs and TRNG}
							\begin{tabular}{|l|l|l|l|l|l|}
									\hline
									$\textbf{Criterian}$ & \textbf{/dev/(u)random} & \textbf{Yarrow}  & \textbf{Fortuna} & \textbf{KCS-PRNG} & \textbf{TRNG} \\ \hline \hline
									\begin{tabular}[c]{@{}l@{}}Hard problem used \end{tabular} &  \begin{tabular}[c]{@{}l@{}}ChaCha20 \\Stream cipher \end{tabular} &	 \begin{tabular}[c]{@{}l@{}}3DES \end{tabular} &  \begin{tabular}[c]{@{}l@{}}AES128 in\\ counter mode \end{tabular} & ECDLP  & \begin{tabular}[c]{@{}l@{}}Physical property of \\Oscillator-Rings \end{tabular}\\  \hline
									Hash function &  \begin{tabular}[c]{@{}l@{}}SHA160, MD5 \\ \cite{rock2005pseudorandom} \end{tabular} & SHA160 & SHA256	& SHA256  & \begin{tabular}[c]{@{}l@{}}Not applicable \end{tabular}\\  \hline
									
									\begin{tabular}[c]{@{}l@{}}RNG requirements met\end{tabular} & $R\ref{R1},R\ref{R2}, R\ref{R3}$ & $R\ref{R1},R\ref{R2},R\ref{R3}$ & $R\ref{R1},R\ref{R2},R\ref{R3}$ &  \begin{tabular}[c]{@{}l@{}}$R\ref{R1},R\ref{R2},R\ref{R3}$, \\ $R\ref{R4}$ (Mitigated) \end{tabular} & \begin{tabular}[c]{@{}l@{}}$R\ref{R1},R\ref{R2},R\ref{R3}$, $R\ref{R4}$ \end{tabular}\\	  \hline
									
									\begin{tabular}[c]{@{}l@{}}Unblocked supply of random\\ bits \end{tabular}& No  & No & Yes	& Yes  & Yes\\ \hline
									
									Correlation Test & *& *& * & \begin{tabular}[c]{@{}l@{}}Passed (serial correlation of \\0.000034) \end{tabular} & Passed\\  \hline
									\begin{tabular}[c]{@{}l@{}}Per bit entropy rate \end{tabular}& *& *& * &	0.99999975 & \begin{tabular}[c]{@{}l@{}}0.9993 \end{tabular}\\  \hline
									
									\begin{tabular}[c]{@{}l@{}}Linear complexity $LC(x)$ \end{tabular} & *& *& * &  \begin{tabular}[c]{@{}l@{}}$15670^{6237} < LC(x) \leq 15670^{6238}$ \end{tabular} & \begin{tabular}[c]{@{}l@{}}Not applicable \end{tabular}\\  \hline
									
									Period	& * &	* &  \begin{tabular}[c]{@{}l@{}}$2^{128}$ in \\ single call \\ \cite{mcevoy2006fortuna} \end{tabular} &  \begin{tabular}[c]{@{}l@{}}$[N_1 \times 2^{401}, (N_1 + N_2) \times 2^{401}]$ \end{tabular} & \begin{tabular}[c]{@{}l@{}}Infinite \end{tabular}\\  \hline
									
									Key space	& * &	* & *	& $[2^{529}, \infty)$ & \begin{tabular}[c]{@{}l@{}}Infinite \end{tabular}\\ \hline
									
									Throughput & 8-12Kbps \cite{rock2005pseudorandom} &  \begin{tabular}[c]{@{}l@{}}	No results \\\cite{rock2005pseudorandom} \end{tabular} & 7.2 Mbps	& 2.5 Mbps & \begin{tabular}[c]{@{}l@{}}6 Mbps on Xilinx \\Spartan-3A FPGA \end{tabular}  \\ \hline
									
									\begin{tabular}[c]{@{}l@{}}Statistical tests passed \end{tabular} &  \begin{tabular}[c]{@{}l@{}}Diehard \cite{rock2005pseudorandom} \end{tabular} &  \begin{tabular}[c]{@{}l@{}}Not available \\ \cite{rock2005pseudorandom} \end{tabular} &  \begin{tabular}[c]{@{}l@{}}Diehard \cite{mcevoy2006fortuna} \end{tabular} &  \begin{tabular}[c]{@{}l@{}}NIST, Diehard, TestU01 \end{tabular} & \begin{tabular}[c]{@{}l@{}}NIST \end{tabular}  \\ \hline
									
									NIST proportion obtained	& * &	* & *	& 0.9896 & \begin{tabular}[c]{@{}l@{}}0.987 \end{tabular}\\ \hline
									
									Restart/Non-reproducibility Test	& * &	* & * & Passed & Passed\\ \hline
								\end{tabular}
								\label{compCSPRNGs1}
							} \newline \newline
							\scriptsize{*No reference available}
						\end{table}
					\end{center}
				\end{landscape}
				\endgroup

	\section{Conclusion} \label{sec11}
				The operating system kernel demands a high quality CSPRNG for its randomness requirements. A novel CSPRNG called KCS-PRNG is presented in this paper which exhibits qualities of a CSPRNG as well as of a TRNG (i.e., it also includes non-reproducibility of the generated random bitstreams) for use in kernel and in various other cryptographic applications. The combination of clock-controlled LFSRs as a nonlinear sequence generator and two non-standard and trusted elliptic curves is proven to be an excellent choice of designing a CSPRNG.~An extensive security analysis of the proposed KCS-PRNG was performed which proved that the proposed generator is resistant to important attacks like Berlekamp-Massey attacks, brute force attacks, next-bit tests, state compromise extension attacks and correlation attacks on the proposed generator.~The proposed design of the KCS-PRNG allows periodic change of elliptic curves in the elliptic curve look-up table maintained by the generator to mitigate the gap of the security property $R\ref{R4}$ i.e., `non-reproducibility' requirement to a practical extent for the first time in the literature. The use of elliptic curves from its look-up table makes the proposed KCS-PRNG customizable than the current popular kernel CSPRNGs like /dev/random, Yarrow and Fortuna whose designs are based on block ciphers like Triple DES and AES respectively.~Hence, it is inferred that the proposed KCS-PRNG qualifies as a competent CSPRNG for adoption in the kernel applications.

				\begin{acknowledgements}
					The authors thank Society for Electronic Transactions and Security (SETS), Chennai for providing the research opportunity to carry out this proposed work. The authors show their deepest gratitude to Dr. P. V. Ananda Mohan and Dr. Reshmi T. R. for their inputs and anonymous reviewers for their review and Mr. T. Santhosh Kumar for help in experimentation. Authors also thank to Mr. Ritesh Dhote, Mr. Aditya Saha, Ms. Sonal Priya Kamal and Ms. Diya V. A. for help in final formatting.
				\end{acknowledgements}

				
\end{document}